\documentclass[11pt,twoside]{article}
\usepackage{asp2010}

\resetcounters

\begin{document}

\title{Open Science Project in White Dwarf Research}
\author{Tommi Vornanen,$^1$
\affil{$^1$Department of Physics and Astronomy, University of Turku, Turku, Finland}}
%\affil{$^2$Institution Full Address for Author2}
%\affil{$^3$Institution Full Address for Author3}}

\begin{abstract}
I will propose a new way of advancing white dwarf research. Open science is a method of doing research that lets everyone who has something to say about the subject take part in the problem solving process.

Already now, the amount of information we gather from observations, theory and modelling is too vast for any one individual to comprehend and turn into knowledge. And the amount of information just keeps growing in the future. A platform that promotes sharing of thoughts and ideas allows us to pool our collective knowledge of white dwarfs and get a clear picture of our research field. It will also make it possible for researchers in fields closely related to ours (AGB stars, planetary nebulae etc.) to join the scientific discourse. 
	
In the first stage this project would allow us to summarize what we know and what we don't, and what we should search for next. Later, it could grow into a large collaboration that would have the impact to, for example, suggest instrument requirements for future telescopes to satisfy the needs of the white dwarf community, or propose large surveys. 
	
A simple implementation would be a wiki page for collecting knowledge combined with a forum for more extensive discussions. These would be simple and cheap to maintain. A large community effort on the whole would be needed for the project to succeed, but individual workload should stay at a low level. 

\end{abstract}

\section{Introduction}
In the past few decades astrophysical research has evolved into a data intensive field of research where the amount of information is overwhelming. Current surveys like the Sloan Digital Sky Survey (SDSS) discover vast numbers of new stars and other objects in all categories and this amount of new objects to study will only increase in the future with missions like Gaia and Pan-STARRS. When these programmes start to release their data products, the information we researchers have to digest is going to increase to terrifying quantities.

Although white dwarfs (WDs) will comprise only a small portion of all the objects, SDSS has shown with its 20000+ WDs that the amount of WDs that will be discovered will be in the hundreds of thousands. It is obvious that no single researcher can hold all the information we get from these stars and form a "bigger picture" about them.

Fortunately advancements in technology have made it possible for large numbers of scientists to work together and share the burden. We are still almost as limited by large distances and slow means of travel as our predecessors 200 years ago, but electronic communication has made modern correspondence much faster than carrier pigeons or messengers riding horses. Still, our methods of keeping in touch with other scientists and their work are essentially the same as those used two centuries ago. They are much faster, but nothing fundamental has changed in the way we work together, despite all the technological advancements at our disposal. We do research at our own laboratory, sometimes travel to another laboratory or telescope, write an article and have it published in a scientific journal. And occasionally we gather together to report on our findings in a conference and socialize with our colleagues.

Nowadays, technology allows us to find new ways of working together to achieve our goals and open science is an umbrella term for a way of doing science that emphasizes cooperation and sharing of knowledge. In essence it means that a group of researchers solve a problem together openly in a manner that allows anyone to take part in the process. As usually the total is more than the sum of its individual parts, an open collaboration can achieve something the individuals by themselves could not have done.

%The obvious benefit of this method is of course that someone might notice something someone else has not. Or notice it faster. This leads to a more efficient, and usually more inspiring, working environment.

\section{Ways of Doing Open Science}
Would open science have something to give to the white dwarf community? Of course! Any field of research, or in general, any endeavour in life benefits from improved methodology. And as long as success in science is measured in the number of scientific papers a researcher produces, we have to make sure that any new method helps us do that. Fortunately, open science projects have been tried before in various fields so we have some precedents to guide us in choosing those methods that actually are better than the ones we use at the moment.

Sharing of data is one of the most important and common ways of opening scientific research. Data archives are already commonplace in astronomy, and in other fields as well. Observatories are in the habit of archiving data and providing access to it for anyone who wishes to do so. There is usually a one year proprietary period during which only the people who took the data are allowed to use it, but after that anyone can download the data and do what they will with it. This is an excellent arrangement and makes the most of precious telescope time. Sometimes the data is used for re-analysis, or sometimes a CCD frame might include something in which the original investigators had no interest. A couple of examples of archival data in use are provided by \citet{tuo09} who discovered a second planet around a star using a different method on the same dataset as the original investigators and \citet{lod12} who cross-referenced several archives to find new subdwarf candidates and then verified some of these using follow-up observations. The problem with using archival data is that you have to know what you are looking for, because search functionality is often quite limited. 

Big astronomical organizations have embraced sharing of data in recent years and are developing various "virtual observatory" projects that aim for the development of various tools that allow a more efficient use of data archives.

Another popular way of cooperation within a research community are wiki pages for collecting the knowledge in a certain field in one place for all to use. This is a brilliant idea and usually everyone supports these kinds of efforts. The harsh truth is then revealed when it is time to start contributing to such a wiki. Hardly anyone in interested in doing so. And the reason usually is that a researcher's time is better spent in writing peer reviewed articles since these are usually the defining factor when choosing a person for an academic position.

One example of a somewhat failed wiki page is the Qwiki.\footnote{URL: http://qwiki.stanford.edu} It is a wiki page dedicated to quantum computing and other advanced quantum related topics.

One good way of increasing the rate of scientific discovery is increasing communication between researchers. A good example of this are scientific conferences where people from the same, or closely related, field meet and discuss their research. This often leads to new ideas and collaborations. But these days we don't have to wait for the next conference or spend money and time travelling there. Technology gives us the means to have these discussions online wherever we are and whenever we have the time for them. 

The power of online discussions is clearly demonstrated by the Polymath Project that started in January 2009 and in a few weeks solved a difficult mathematical problem using a blog.\footnote{URL: http://gowers.wordpress.com/2009/01/27/is-massively-collaborative-mathematics-possible/ and following posts.} The success of that project has led to a multitude of mathematical blogs using the same idea to produce proofs and derivations to various problems.\footnote{See http://michaelnielsen.org/polymath1/ for a collection of polymath projects.} 

The secret behind these successes is that something large and significant comes from small contributions. Basically, a discussion platform like a blog or a forum allows the researchers to connect their brains to form a kind of a super-brain, which makes it possible to solve very difficult problems very easily. One might compare the procedure to connecting a large number of computers together to form a very powerful computing grid.

The reason why Qwiki was a failure and the Polymath Project was successful was pointed out by Michael Nielsen in his TED talk (Technology, Entertainment, Design).\footnote{URL: http://www.ted.com/talks/michael\_nielsen\_open\_science\_now.html } The Polymath Project resulted in peer reviewed articles, although they were written under a pseudonym, the Qwiki did not. Most scientists agree that projects like the Qwiki are very valuable and useful but don't want to spend their precious working hours in contributing text.

\section{Implementation in White Dwarf Research}
Armed with the insights from the previous section and keeping in mind the lessons learned from previous failures and successes, we can take a look at what tools we need to make white dwarf research more open.

A simple wiki page is clearly not a good idea. Our community is very small and studies have shown that only 1 in 5000 users of a wiki page actually contribute text.\footnote{URL: http://strategy.wikimedia.org/wiki/Wikimedia\_users} In our field that would probably mean about 0.1 researchers. A wiki page would without doubt be extremely useful for everyone, especially for the young researchers. Maybe in the future we will have one, but building it up will be a slow task.

We definitely need a discussion platform. Polymath projects use blogs and they are great for discussions on a single topic. However, in the white dwarf community we have numerous topics to discuss and we need to have several threads to talk about them all. A forum would be an ideal platform for the needs of our community as it allows several different discussions in separate threads simultaneously. One important thing to remember, though, is \LaTeX \, compatibility so formulas can be written with less effort, in case things get out of hand and we find ourselves using mathematics.

Although a discussion platform is the most important thing in a project like this, there are other tools that the community might find useful. For example, from the comments I received during the conference, I found out that people in our community would like to have some sort of a paper rating and commenting functionality. Something like the user reviews on items in a webstore. An easy implementation might be the user rating and comment part of a webstore application and the item would be just a link to the article in arXiv. This tool would make it faster to find relevant information about any topic within WD research and also to make the amount of new articles one has to read smaller.

Apart from a forum we could also benefit from posting online our ideas about what to research. This could work as a sort of sanity check on the ideas as well as a way to form collaborations. It would also make life a little easier for those researchers who are trying to find their next project. Most of us get ideas that we do not have time to explore in any way, but that does not mean those ideas should be kept buried in a desk drawer.

It would also useful to be able to share data, or at least metadata about our observations, that we possess. Not all data gets published and most observers have so much data that they do not have time to study all of it thoroughly. Also, archives are only starting to provide reduced data products and it does not always make sense to start the data reduction from scratch. And this kind of sharing could be extended to future projects as well. That way we will not accidentally observe something twice. And it would also make it possible to synchronize complementary observations of an object.

There exists a database of information about all white dwarfs called the Villanova White Dwarf Catalogue \citep{mcc99}\footnote{URL: http://www.astronomy.villanova.edu/WDCatalog/index.html}, but it lacks proper search functionality. It is not worth the effort to re-invent the wheel, but perhaps the community could think of ways to improve the catalogue to make it more useful and incorporate it into the open science project.

Here were a few examples of what kind of things an open science project website could include. But the great thing about this kind of a project is that we can add things, if the community finds them useful. I have started a blog for planning the project and I invite everyone to take part in discussions there\footnote{URL: http://wdopenscience.blogspot.com/} and on the final open science website.

%\acknowledgements 

\bibliography{vornanen}

\end{document}